\documentclass[a4paper]{spie}  
\def\lesssim{\mathrel{\hbox{\rlap{\hbox{\lower4pt\hbox{$\sim$}}}\hbox{$<$}}}}
\def\sax{{\em BeppoSAX\/}} 
\def\fermi{{\em Fermi\/}}
\def\swift{{\em Swift\/}}  
\def\xte{{\em Rossi--XTE\/}}
\def\integral{{\em INTEGRAL\/}}
\def\nustar{{\em NuSTAR\/}}

\usepackage[]{graphicx}
\title{Scientific prospects in soft gamma-ray astronomy enabled by the LAUE project} 
\author{F.~Frontera\supit{a,b}, E.~Virgilli\supit{a}, V.~Valsan\supit{a,c}, V.~Liccardo\supit{a,c}, V.~Carassiti\supit{d}, E.~Caroli\supit{b},  F.~Cassese\supit{f},  C.~Ferrari\supit{e}, V.~Guidi\supit{a}, S.~Mottini\supit{h}, M.~Pecora\supit{g}, B.~Negri\supit{k}, L.~Recanatesi\supit{f}, L.~Amati\supit{b}, 
N.~Auricchio\supit{b},   L.~Bassani\supit{b}, R.~Campana\supit{b}, R.~Farinelli\supit{a}, C.~Guidorzi\supit{a},  C.~Labanti\supit{b}, R.~Landi\supit{b}, A.~Malizia\supit{b}, M.~Orlandini\supit{b}, P.~Rosati\supit{a}, V.~Sguera\supit{b}, J.~Stephen\supit{b}, L.~Titarchuk\supit{a}  
\skiplinehalf
\supit{a} \small\textit{Department of Physics and Earth Science, University of Ferrara - Via Saragat, 1, 44100 Ferrara - Italy};\\
\supit{b} \small\textit{INAF-IASF Bologna, via P.~Gobetti 101, Bologna - Italy};\\
\supit{c}\textit{Universit\'e de Nice Sophia Antipolis, Nice, Cedex 2, Grand Chateau Parc Valrose - France;}\\
\supit{d} \small\textit{INFN, Sezione di Ferrara, via Saragat 1, 44100 Ferrara - Italy};\\
\supit{e} \small\textit{CNR, IMEM, Parco Area delle Scienze 37/A - 43124 Parma - Italy};\\
\supit{f} \small\textit{DTM, Modena, Via Tacito, I-41100 Modena - Italy};\\
\supit{g} \small\textit{Thales Alenia Space--Italy, Milan - Italy};\\
\supit{h} \small\textit{Thales Alenia Space--Italy, Turin - Italy};\\
\supit{k} \small\textit{ASI, Agenzia Spaziale Italiana, Viale Liegi 26, I-00198 Roma - Italy}.
}

\authorinfo{E-mail to: frontera@fe.infn.it}
\begin{document} 
\maketitle 

\begin{abstract}
This paper summarizes the development of a successful project, LAUE, supported by the Italian Space Agency (ASI) and devoted to the development of long focal length (up to 100~m) Laue lenses  for hard X--/soft gamma--ray astronomy 
(80-600 keV). The apparatus is ready and the assembling of a prototype lens petal is ongoing. The great achievement of this project is the use of bent crystals. From measurements obtained on single crystals and from  simulations, we have estimated the expected Point Spread Function and thus the sensitivity of a lens made of petals. The expected sensitivity is
a few $\times10^{-8}$~photons~cm$^{-2}$~s$^{-1}$~keV$^{-1}$. We discuss a number of  open astrophysical questions that can settled with such an instrument aboard a free-flying satellite.
\end{abstract}


\section{INTRODUCTION}
\label{sec:intro}

Hard X-/soft gamma-ray astronomy is a crucial  window for the study of the most energetic and violent events in the Universe.  
With the advent of the NASA \nustar\ mission \cite{Harrison13}, having on board two focusing telescopes operating in the 3--79 keV band, very sensitive ($10^{-8}$~photons~cm$^{-2}$~s$^{-1}$~keV$^{-1}$)) studies of the hard X--ray sky  finally are becoming possible. \nustar\ has already performed sudies of single sources (e.g., the hard X--ray emission from the blazar MKN 421 \cite{Balakovic13}, the nuclear emission from the local star-forming galaxy NGC\,253 \cite{lehmer13}, and the hard X--ray emission accreting pulsar 
GS\,0834$-$530 \cite{Miyasaka13}), studies of classes of sources, like the broad absorption line quasars PG\,1004$+$130 and PG\,1700$+$518 \cite{luo13}, and extragalactic surveys of AGNs to resolve the Cosmic X--ray Background (CXB) at the peak of the $E F(E)$ spectrum \cite{Alexander13}. 

A significant part of the \nustar\ core program time will be devoted to sensitive studies of extragalactic and Galactic surveys, which will supplement the extended surveys beyond 20 keV already performed \cite{Bird10,Cusumano10} or planned to be continued with the ESA INTEGRAL observatory \cite{Winkler03}, 
and with the NASA \swift\ satellite \cite{Gehrels04}.

%
Evidence of extended matter-antimatter annihilation
emission (at 511 keV) from the Galactic  Center \cite{Weidenspointner08} and of Galactic
nucleosynthesis processes \cite{Weidenspointner08,Diehl06} has been found. Furthermore, polarization of high energy photons ($>$ 400 keV) emitted from a strong source like the Cygnus X-1 has been clearly measured \cite{Laurent11}.

However, in order to take full advantage of these results, a new generation of focusing telescopes which extend the energy band up to several hundreds of keV is needed. In spite of its high sensitivity, \nustar\ optics are based on multilayer reflectivity and therefore are limited in the energy passband. 
A two--order of magnitude increase in sensitivity  and angular resolution will lead to a leap forward in our understanding of
the (especially non--thermal) processes occurring in the hard X--ray Universe, in similar way to the soft X--ray optics in the seventies.

Laue lenses, based on diffraction from crystals in a transmission configuration,
offer the best technical solution to the implementation of a focusing telescope that can extend the energy band beyond 80 keV. Indeed, the adoption of grazing incidence mirrors, like those used for the telescopes aboard, e.g., {\em Chandra} or {\em XMM-Newton}, or of multilayer mirrors, like those used for the \nustar\ telescopes, would require hundred meter focal lengths to extend the energy band up to several hundreds of keV. The advantage of Laue lenses is that we can extend the band up to 600 keV with a focal length of about 20~m, which is still feasible with a single satellite.

The assembling of Laue lenses, which are made from thousands of crystals, is not a simple task, however we report here the most recent progress of the LAUE project, where we have developed the methodology for an accurate assembling of a lens in a relatively short time. 
With this technique, we are confident that a new window in hard X--/soft--gamma ray astronomy will be opened in the near future, with a big leap in both flux sensitivity and angular resolution.

\section{The LAUE Project}

\subsection{Goals and motivations}

We have already reported on the LAUE project \cite{Virgilli11b,Frontera12,Valsan12} in the past few years. 
It is the follow-up of an initial development project (HAXTEL) in which the adopted lens-assembling technique was found to be suitable only for lenses with short focal--length ($<$10~m) \cite{Frontera08a,Virgilli11a}. The crystals used (flat tiles of $15 \times 15$~mm$^2$ cross section) were made of Cu (111) in mosaic configuration provided by the Institute Laue Langevin (ILL) in Grenoble, France. The lens-assembling required several steps with a resulting cumulative error budget of a few arcmin, that was not acceptable for a space Laue lens with a long focal length (at least about 20 m). An additional problem was the crystal tile production. The only material available in mosaic configuration  was the Copper provided by ILL, that was produced only in limited amounts for neutron ground-experiments, not adequate for building Laue lenses.
Building on this experience, we established the following requirements for the LAUE project:

\begin{itemize}

\item to develop an advanced technology with a cumulative error budget lower than 10 arcsec, leading to a sensitive Laue lens with long focal length (at least 20 m). The goal was to extend the band beyond that of the multilayer telescopes ($>$ 80 keV) up to 600 keV;
\item to solve the still long standing and difficult issue of developing crystals suitable for Laue lenses: explore the use of bent crystals, and their massive production;
\item to build a lens petal made of these newly developed crystals;
\item to perform a feasibility study of a space lens made of petals.
\end{itemize}

Concerning the first item,  based on the experience acquired with HAXTEL, in order to decrease the cumulative error budget, we decided to develop  an apparatus to correctly orient and fix each crystal to the lens frame under the control of a gamma--ray beam.
For the same reason, we also decided to keep the lens petal fixed and to move the gamma--ray beam parallel to the lens axis.

Concerning the second item, the development of bent crystals was considered of key importance, in order to increase the reflection efficency and to obtain a better focusing of the photons.

\subsection{Developed apparatus and its location}

\begin{figure}[!h]
\begin{center}
\includegraphics[width=0.37\textwidth]{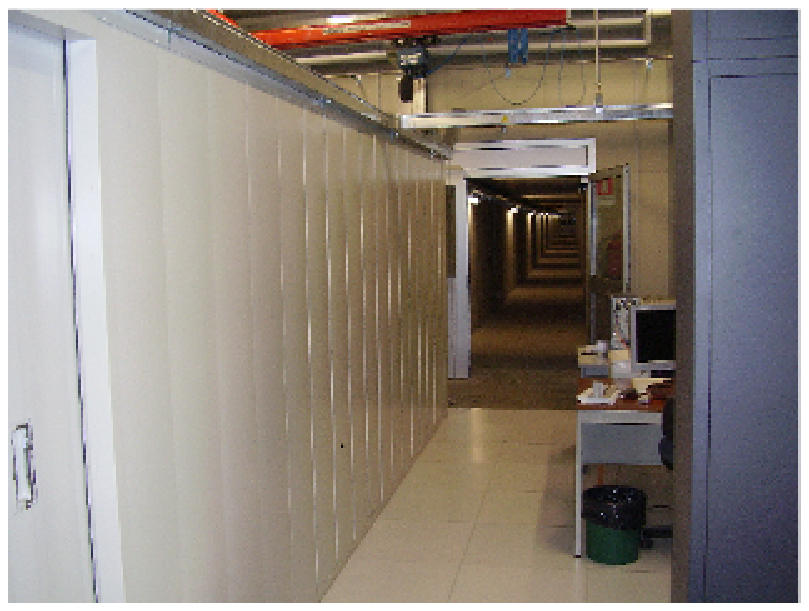}
\includegraphics[width=0.37\textwidth]{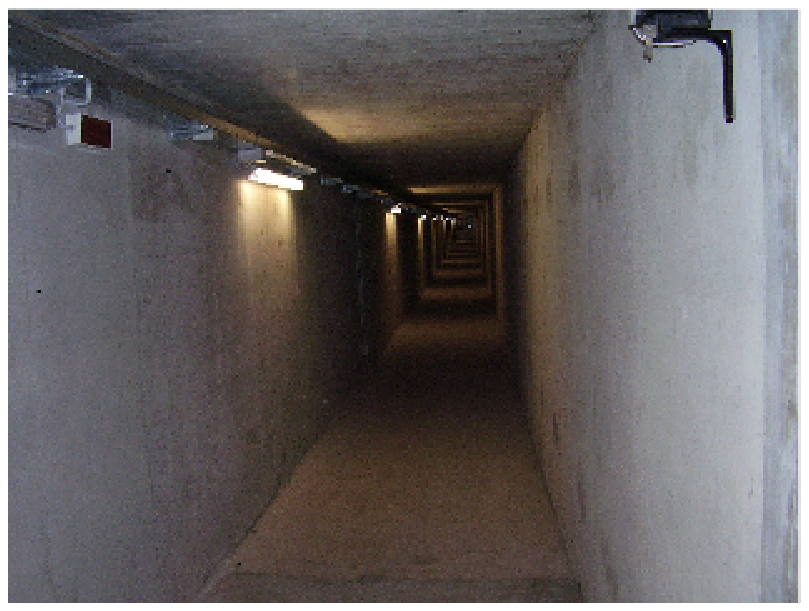}
\end{center}
\caption{{\em Left panel}: View of the entrance in the LARIX tunnel at the Ferrara University. Before the entrance, it is visible the shielded wall of a 12 m X-ray facility. {\em Right panel}: the LARIX tunnel at the beginning of the LAUE project.}
\label{f:tunnel} 
\end{figure}

The first issue to be solved was to have a suitable location for the apparatus to be built. The solution was to use the 100 m long tunnel of the LARIX Laboratory of the University of Ferrara (see Figure~\ref{f:tunnel}), after  realization of the the required infrastructure (shielded doors and walls, clean room of class better than 10$^5$, with humidity and temperature control).

%

\begin{figure}[!h]
\begin{center}
\includegraphics[width=0.37\textwidth]
{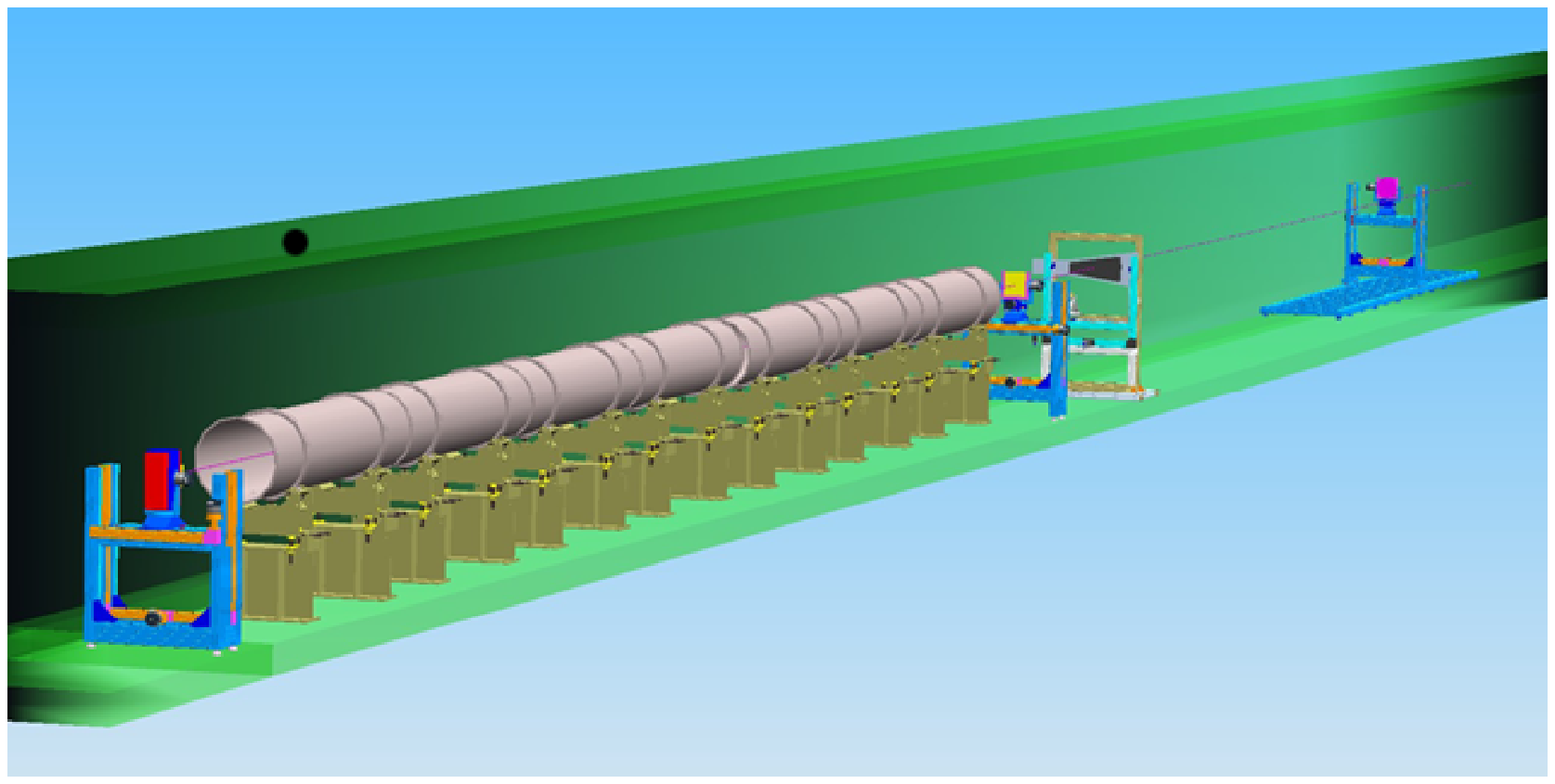}
\includegraphics[width=0.37\textwidth]
{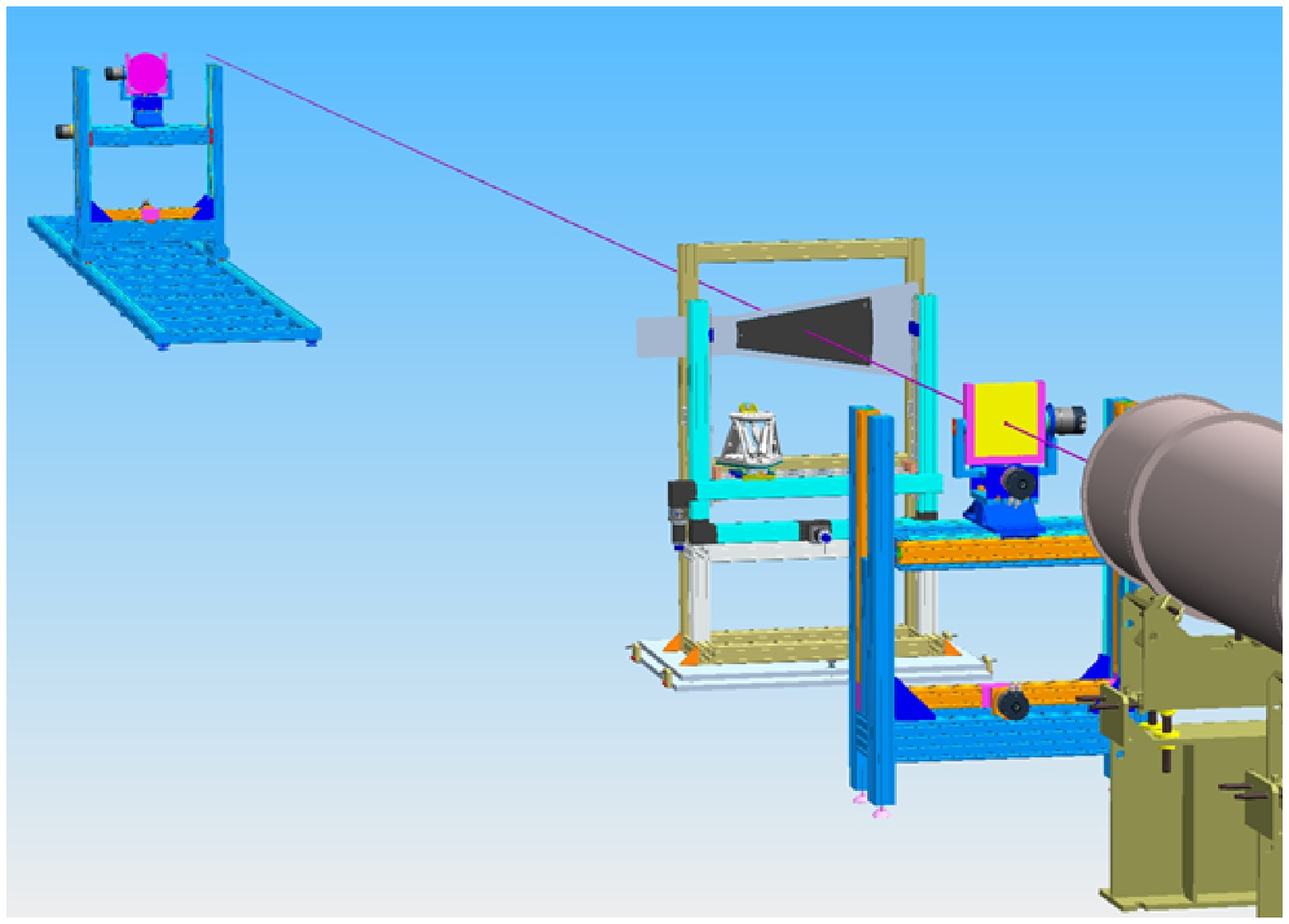}
\end{center}
\caption{{\em Left panel}: Sketch of the LAUE apparatus located in the tunnel of the LARIX facility of the Ferrara University. It shows the X--ray source, the beam line, the lens location, and the focal plane detectors mounted on a rail. {\em Right panel}: zoom on the latter part of the instrumentation chain.}
\label{f:apparatus} 
\end{figure}

The layout of the developed apparatus is shown in Figure~\ref{f:apparatus}, with details reported in another paper (Virgilli et al., this volume).  

%

\begin{figure}[!h]
\begin{center}
\includegraphics[width=0.5\textwidth]{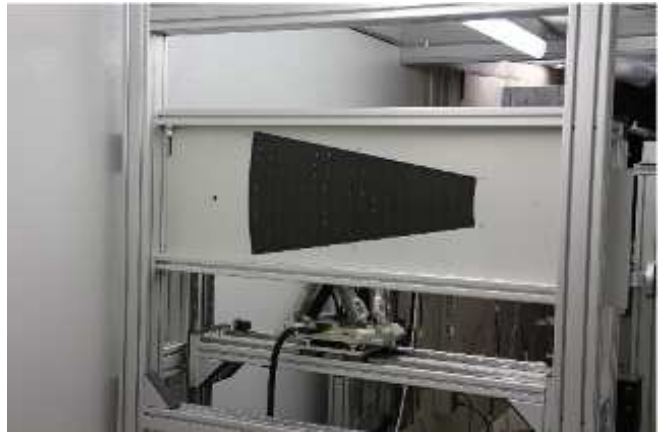}
\end{center}
\caption{Front view of the lens petal. Also shown is the hexapod for the fine positioning of the crystals on the frame.}
\label{f:frame} 
\end{figure}

In short, the apparatus consists of:
\begin{enumerate}
\item A gamma--ray source produced by an X--ray tube with maximum voltage of 320 kV and a very small spot size (0.4~mm diameter) from which the photons are emitted. The source photons are mechanically collimated  to get an output beam of about 20 arcmin. The source can be translated in a plane ($y$,  
 $z$) perpendicular to the beam axis (called $x$ axis) and tilted around axes $z$ (vertical axis) and $y$.
\item A 21~m beamline (60 cm inner diameter) within which the gamma--ray beam travels under vacuum to avoid absorption and scattering interactions. The beamline was initially designed to be 70 m long, but, to contain costs, it was shortened to 21 m.
\item At the exit of the beamline, a final shield of the gamma--ray beam with a square hole in the center, where a slit with the aperture defined by 4 independently movable Tungsten Carbide blades is set up. In this way a pencil beam of the required size could be obtained. Both shield and slit can be translated in the plane (y, z) and rotated around the three axes ($x$, $y$, $z$). 
\item Lens--petal frame, which is held  by a pedestal that can be manually rotated and translated. It has the shape of a circular  sector, 18 degree wide as seen from the lens axis
(see Fig.~\ref{f:frame}).
\item A fine six--axes  motorized robot (hexapod) that allows the correct positioning of each crystal tile on the lens frame under the control of the gamma--ray pencil beam. Once the diffracted photons are focused on the lens focus, the crystal tile is fixed to the lens frame.
\item
Two focal plane detectors (a gamma--ray imager and a spectrometer), one above the other, held by a pedestal. The pedestal can be moved along the $x$ axis using a rail oriented along this axis. For a fixed value of $x$, the detectors can be translated in the ($y$, $z$) plane and rotated around each of the three axes $x$, $y$, and $z$. The gamma-ray imager is based on a CsI(Tl) scintillator, whose fluorescence light is viewed from an array of photodiodes, that provides a position resolution of 200~$\mu$m. The spectrometer instead is based on an Ortec High Purity planar Germanium (HPGe) detector cooled at liquid Nitrogen temperature, with an energy resolution of about 500 eV a 60 keV. Details of this instrumentation are reported in another paper (Caroli et al., this volume). 
\item Both the final slit and the petal frame are located in the already mentioned clean room to guarantee the crystal positioning and fixing to the lens frame under constant cleanliness and environment conditions.
\end{enumerate}   
 
All the movements listed above are motorized and controlled by the console room located outside the tunnel (LARIX A laboratory). For the electrical ground support equipment, see paper by Caroli et al. (this volume).

%

\subsection{Developed crystals}

An intense development activity was performed by the crystallographers involved in the project. On one hand, a technology has been developed to grow crystals of GaAs with a mosaicity of about 30 arcsec \cite{Ferrari12} and to bend them. The bending of the GaAs crystals was achieved by a controlled surface damaging, which introduces defects in a layer of few tens of nanometers in thickness undergoing a highly compressive strain \cite{Buffagni12} (see also Ferrari et al., this volume).
On the other hand, we developed a technology for bending perfect crystals of Si(111) and Ge (111) by surface grooving, also obtaining an internal quasi-mosaicity (QM)of the crystals \cite{Guidi12,Camattari13} (see also Guidi et al., this volume).
 
 On the basis of these developments, for the lens-petal prototype, we decided to use both  bent crystals of QM Ge(111) and bent mosaic crystals of GaAs (220), for comparison with the standard mosaic GaAs crystals. In this way, it was possible to achieve two key goals: a much better focusing (see below) and higher reflectivity. We know that bent crystals do not have the limitation of a maximum diffractivity of 50\%, as it is the case for flat crystals. 

A lens petal prototype, made of 150 bent crystals of each type (300 crystals in total), with a nominal curvature radius of 40~m, a cross-section  of 30$\times 10$~mm$^2$ and thickness of 2~mm, will be built as a final product of the LAUE project. The bent GaAs (220) tiles produced for the petal show a mosaicity between 20 and 25 arcsec, while the bent Ge (111) show a QM of about 4 arcsec.
The actual curvature is measured in the LARIX 12 m long X--ray facility, with preliminary results reported in another paper (Liccardo et al., this volume).

It is the first time that bent crystals have sepecifically been developed and used for a Laue lens. It is also the first time that  a lens petal with a so long focal length  (20~m) is   being assembled.

\subsection{Apparatus alignment}

The alignment of the apparatus (see details in the paper by Virgilli et al, this volume) is crucial for the accurate positioning of the crystal tiles in the lens, thus achieving a fine focusing of the reflected photons. The apparatus alignment was performed in three steps, as descibed below:

\begin{enumerate}
\item Mechanical alignment. 

Given that the beam line is fixed, the frame of the 
lens-petal was  mechanically positioned perpendicularly to the beamline axis (previously optically measured). Its cross section was translated in such a way to be within the internal size of the beamline. Then, the gamma--ray source and the final slit were oriented in such a way that they translate in the plane ($y$, $z$) parallel to the lens--frame cross section and thus perpendicular to the beamline axis.
The focal--plane detectors are positioned at 20 m distance from the lens petal and its pedestal is oriented in such a way that they can be translated along the ($y$, $z$) plane.

\item Optical alignment
 
The optical alignment was performed using the beamline without entrance and exit windows. An optical laser and a beam-splitter were positioned between the end of the beamline and the final slit (see Figure~\ref{f:beam-splitter}). They were then oriented in such a way that the opposite beams travelled horizontally along the $x$ axis. It was therefore possible to determine the zero positions of the gamma-ray source collimator output, of the final slit, and of the reference hole in the lens frame, and the zero positions of the focal plane imager and spectrometer. Thus we determined the lens focus.

%
 \begin{figure}[!h]
\begin{center}
\includegraphics[width=0.5\textwidth]{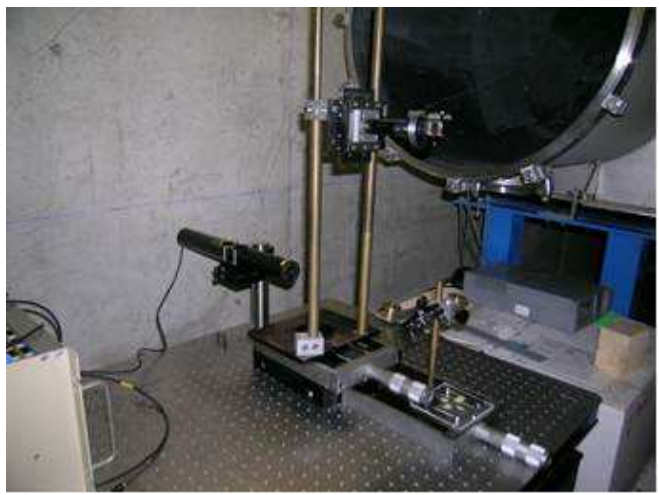}
\end{center}
\caption{Equipment adopted for the optical alignment of the LAUE apparatus. It includes a laser, whose optical beam is made horizontal, a mirror a 45 deg and a beam splitter, that provides two antiparallel horizontal beams.}
\label{f:beam-splitter} 
\end{figure} 

\item Gamma--ray alignment

After the optical alignment, we checked whether the gamma--ray beam in the zero position crossed the slit and the reference cell of the petal frame. This check was performed by using two crosses, made of W wires of 200~$\mu$m diameter, positioned in the center of the slit zero position and in the center of the reference cell of the petal frame. The image of the crosses was detected by the focal plane gamma-ray imager. An additional fine adjustement of the source zero position was required. The final successful result of this alignment procedure can be seen in Fig.~\ref{f:crosses}.

%
 \begin{figure}[!h]
\begin{center}
\includegraphics[width=0.5\textwidth]{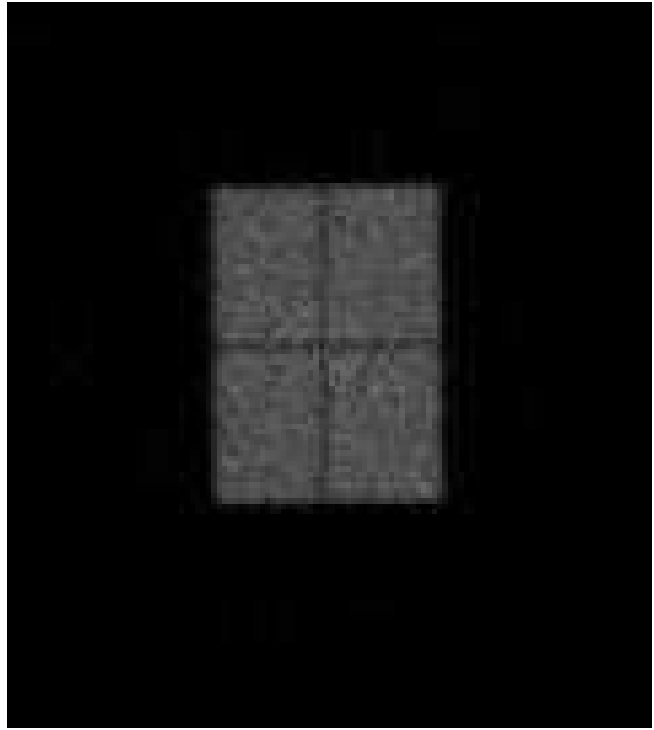}
\end{center}
\caption{Image obtained of the two Tungsten crosses when the apparatus is fully aligned. The imager pixel size is 200~$\mu$m. }
\label{f:crosses} 
\end{figure} 

The monitoring of the beam intensity at the exit of the beamline, before impinging on the slit collimator, is performed using a NaI(Tl) scintillator integrated with a photomultiplier (see Fig.~\ref{f:monitor}).  The count rate in a given energy band is continuously recorded. In this way, the incident beam on the radiated crystal tile can be determined.

%
 \begin{figure}[!h]
\begin{center}
\includegraphics[width=0.5\textwidth]{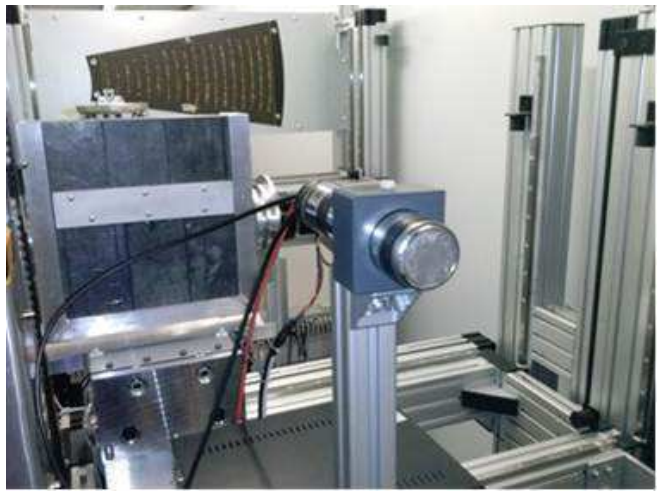}
\end{center}
\caption{The monitor of the incident gamma--ray beam.}
\label{f:monitor} 
\end{figure} 

\end{enumerate} 

\section{Petal focusing capability}

\begin{figure}[!h]
\begin{center}
\includegraphics[width=0.37\textwidth]
{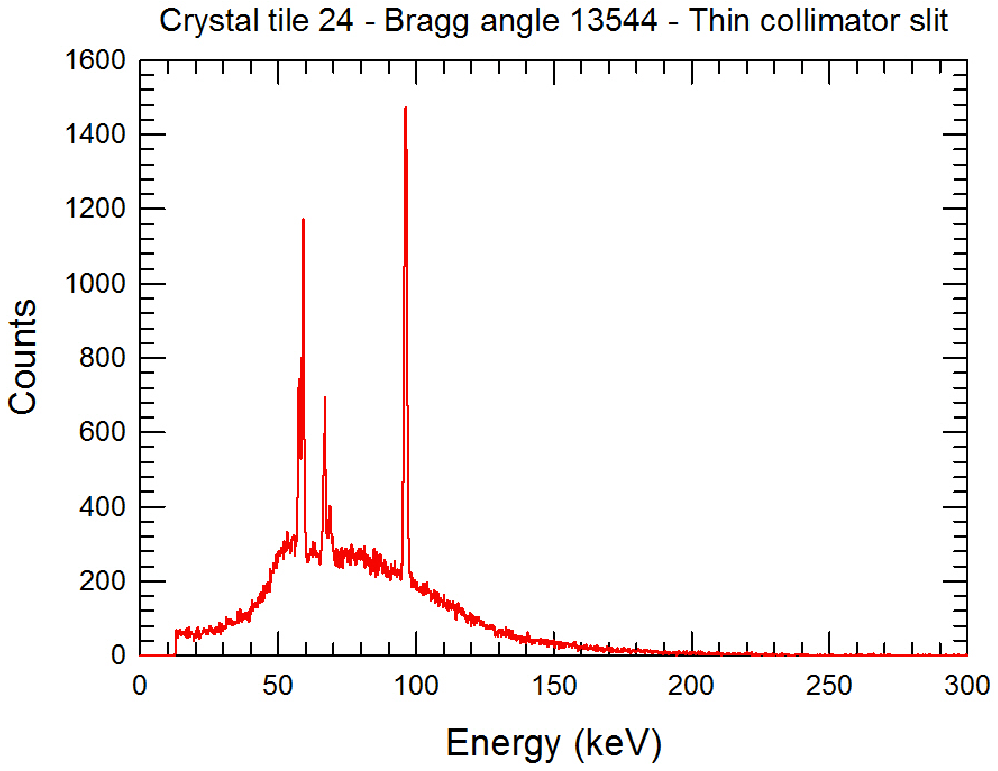}
\includegraphics[width=0.37\textwidth]
{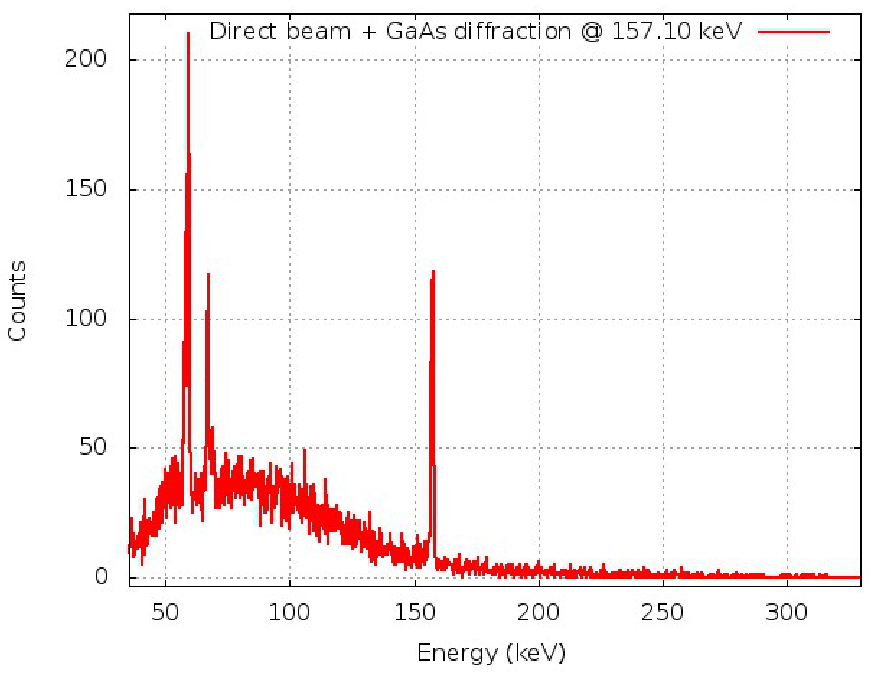}
\end{center}
\caption{{\em Left panel}: Spectrum observed in 300 s with Ge (111) diffraction configuration. Slit size: 0.5 mm along the radial direction and 6 mm in normal direction.  The expected diffracted energy peak is 100 keV. The other peaks and the continuum spectrum is due to the background observed in the focus position. {\em Right panel}: Spectrum observed in 50 s with GaAs (220) in diffraction configuration. Slit size: 0.5 mm along the radial direction and 6 mm in the normal direction.  The expected diffracted energy peak is 157.1 keV. The other peaks and the continuum spectrum are due to the background observed in the focus position. }
\label{f:diffr} 
\end{figure}

Once the apparatus was successfully aligned, a focusing test of crystal samples was performed with both the spectrometer and imager, in order to verify the focusing at the 20 m focal distance.
Both spectra and images were obtained.
Using the HPGe spectrometer in the lens focus, we measured the focused photon spectrum, finding a small deviation (less than 1 keV) from that expected. The correct value of the centroid energy was easily recovered by using the hexapod. The measured spectra of the diffracted beam by Ge (111) and GaAs (220) are shown in Fig.~\ref{f:diffr}.

%

\begin{figure}[!h]
\begin{center}
\includegraphics[width=0.5\textwidth]{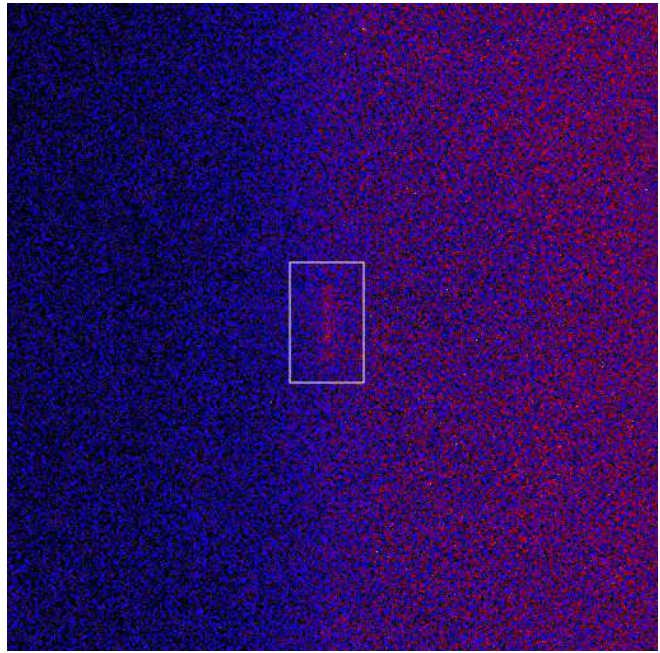}
\end{center}
\caption{Image of the reflected gamma-ray beam (157.1 keV) trough the GaAs(220) bent crystal. Slit aperture size: 4 mm  along the radial direction and 6 mm in the orthogonal direction. The focusing effect along the radial direction is apparent.}
\label{f:GaAs-image} 
\end{figure}

One of the first reflected images by the bent GaAs (220) irradiated crystal in the petal focus (20~m focal length) is shown in  Fig.~\ref{f:GaAs-image}.

For further tests which were performed to check the correcteness of the apparatus alignment,  see Virgilli et al. (this volume).

%
%

\section{Expected sensitivity based on bent crystals of Ge(111) or GaAs(220)}

\begin{figure}
\begin{center}
\includegraphics[width=0.4\textwidth]{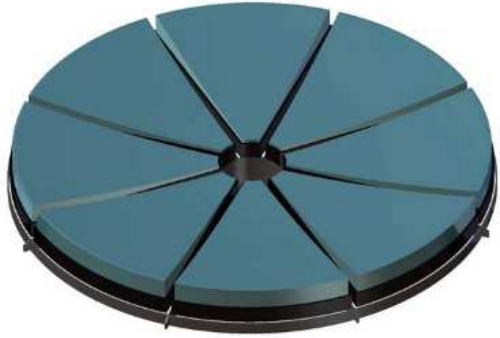}
\end{center}
\caption{A space lens made of petals resulted from the feasibility study performed by TAS-I within the LAUE project.}
\label{f:space-lens}
\end{figure}

The results outlined above show that the developed apparatus has the required performance to allow a lens made of petals to be built. 
Within the LAUE project a feasibility study of  a lens made of petals has been undertaken by Thales-Alenia Space Italy (TAS-I), branch of Turin, with positive results. 
The resulting model of such a study is shown in Fig.~\ref{f:space-lens}.

%
%

\begin{figure}
\begin{center}
\includegraphics[width=0.37\textwidth]{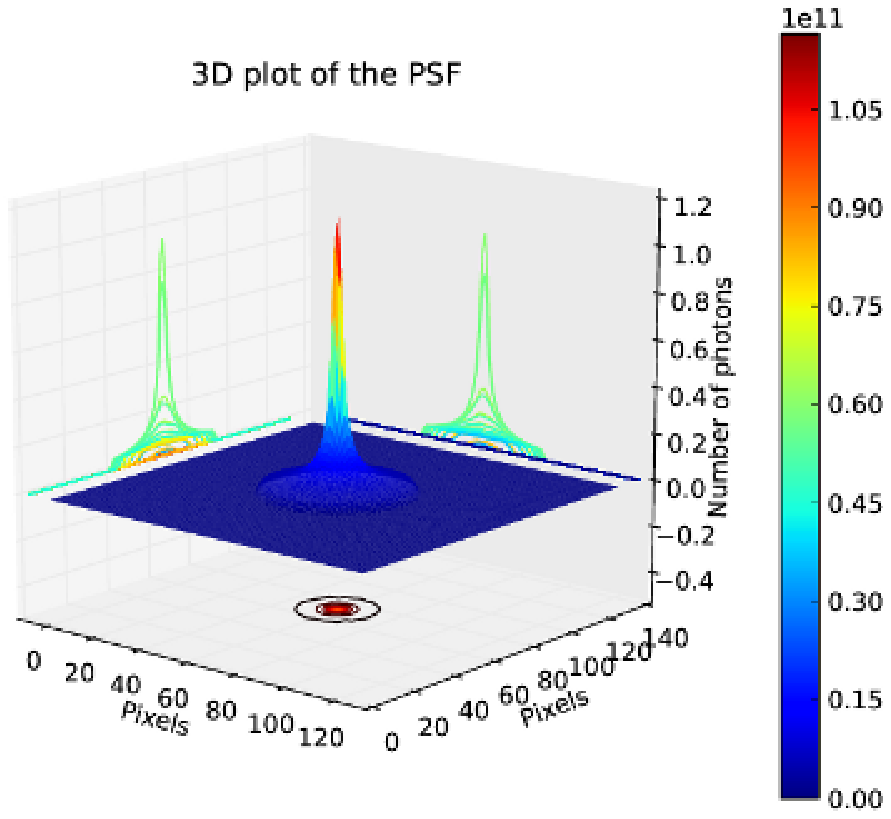}
\includegraphics[width=0.37\textwidth]{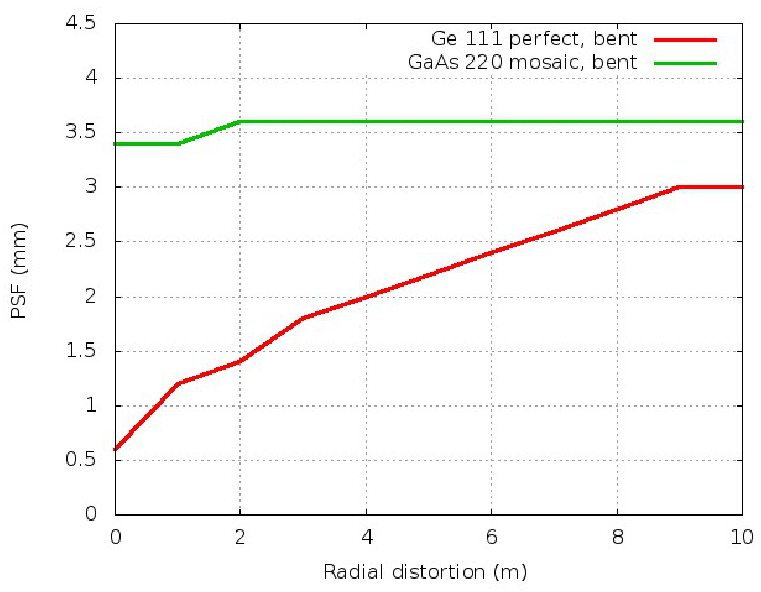}
\end{center}
\caption{{\em Left panel}: three-dimensional PSF of a lens with 20 m focal length, made of bent crystal tiles of Ge (111) (pixel size of 200~$\mu$m).
{\em Right panel}: Dependence of the PSF fwhm on radial distortion.}
\label{f:psf}
\end{figure}

From preliminary results and from simulation studies, we found that the measured reflectivity is close to the expectation, and the Point Spread Function (PSF) of a lens made of petals of bent crystals of either Ge(111) or GaAs(220) is similar to the one in our prototype. We also estimated the effect of radial distortion on the PSF, i.e., deviation of the curvature radius from the nominal of 40 m, on the Full Width at Half Maximum (FWHM) of the PSF. The expected PSF in the case of Ge(111) and the FWHM behavior for both lens types as a function of the radial distortion of the crystal tiles are shown in 
Fig.~\ref{f:psf}.

%
%

\begin{figure}
\begin{center}
\includegraphics[width=0.37\textwidth]{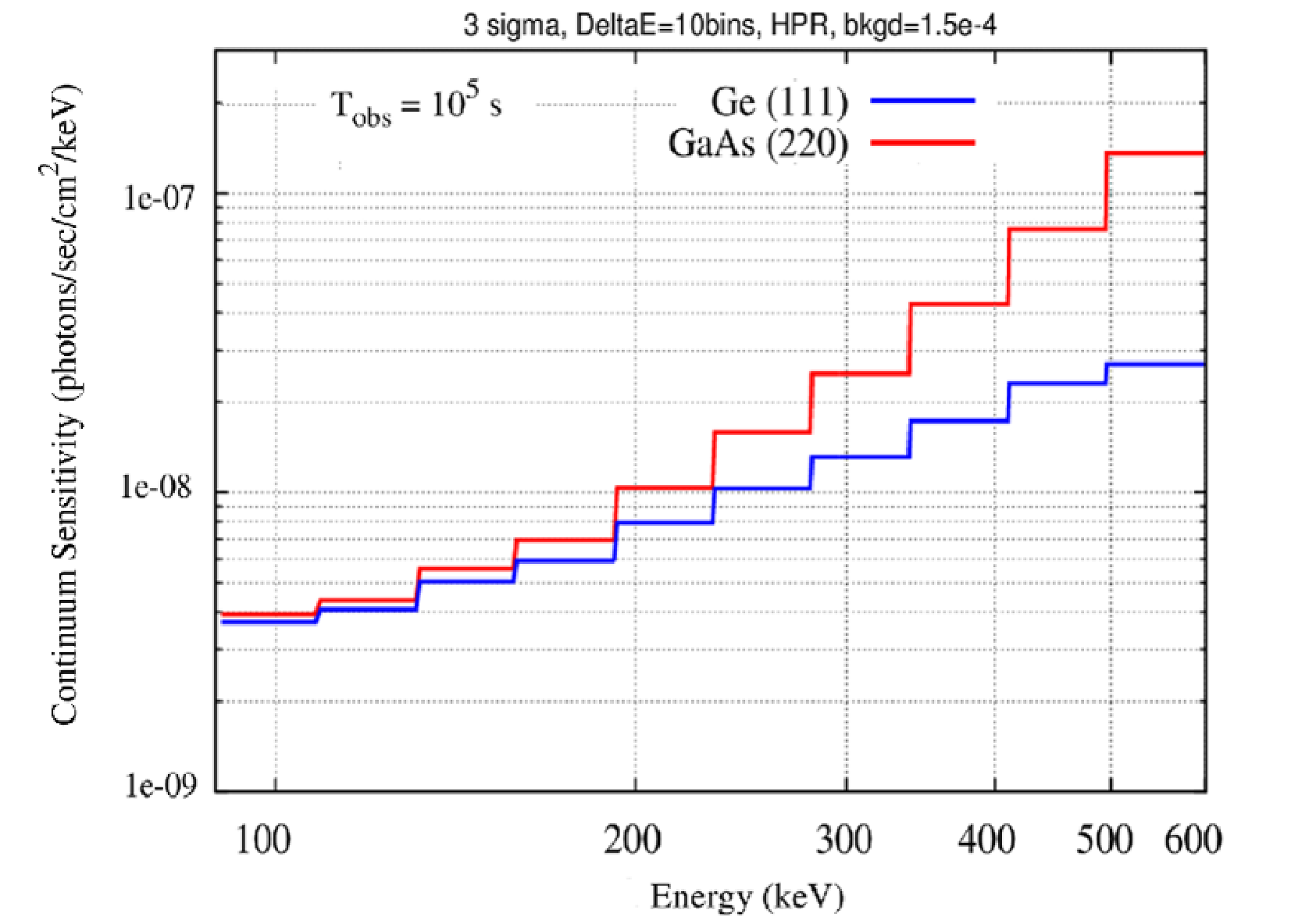}
\includegraphics[angle=0,width=0.37\textwidth]{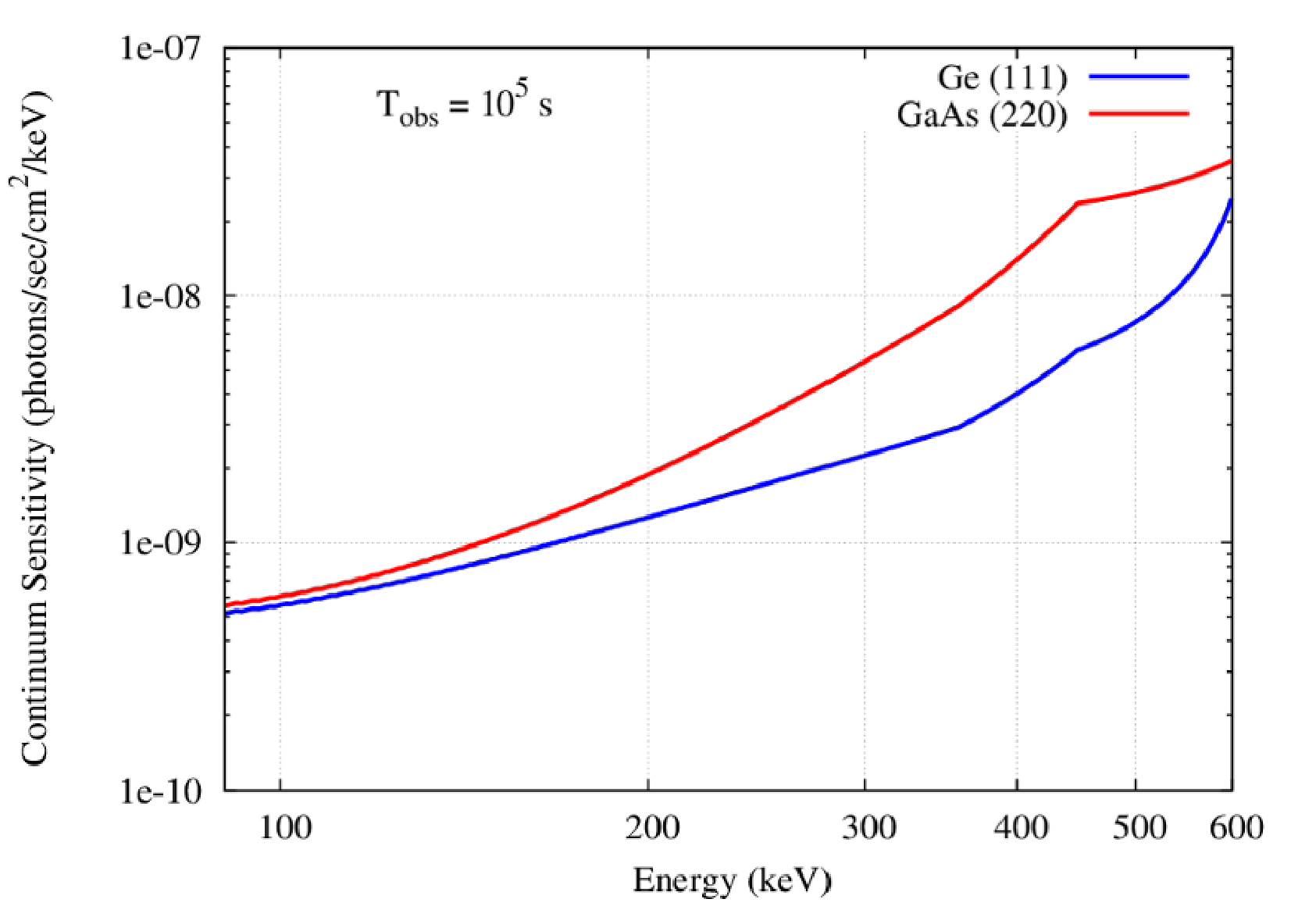}
\end{center}
\caption{Expected on--axis sensitivity, in $10^5$~s, to continuum intensity from a 20~m focal length Laue lens. Image photons are extracted from the Half Power Radius (HPR), background level is 
$1.5\times 10^{-4}$~counts~cm$^{-2}$~s$^{-1}$~keV$^{-1}$, the detection efficiency is 0.9, 3$\sigma$.
{\em Left panel}: Sensitivity in 10 logarithmic energy bins. 
{\em Right panel}: Sensitivity in energy intervals  
$\Delta E = E/2$.}
\label{f:cont-sens}
\end{figure}

Assuming no radial distortion, we have evaluated the expected sensitivity of a lens with 20~m focal length made of bent crystals of either Ge(111) or GaAs(220), that covers the energy band from 90 keV to 600 keV. The outer diameter of the lens is about 168~cm in the case of Ge(111) and 276~cm in the other case.
The result is shown in Fig.~\ref{f:cont-sens}, in which the following assumptions were made: image photons extracted from the Half Power Radius (HPR), background level of 
$1.5\times 10^{-4}$~counts~cm$^{-2}$~s$^{-1}$~keV$^{-1}$, dectection efficiency of 0.9, 3$\sigma$ confidence level.

%
%

\begin{figure}
\begin{center}
\includegraphics[angle=0,width=0.4\textwidth]{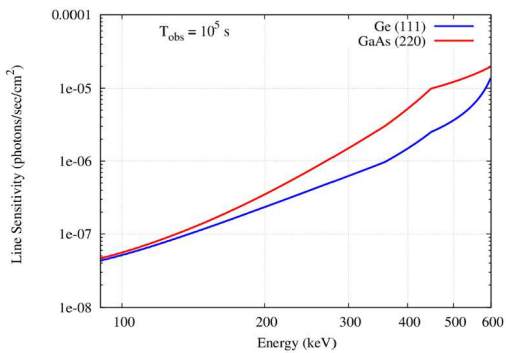}
\end{center}
\caption{Expected on--axis sensitivity in $10^5$~s to emission lines from a 20 m focal length Laue lens. The assumptions are those adopted for the continuum sensitivity. The assumed continuum intensity under the line is the continuum sensitivity with $\Delta E = E/2$.}
\label{f:line-sens}
\end{figure}

We have also derived the expected sensitivity of such a lens to narrow Gaussian lines ($FWHM \approx 1.5$~keV), using the same parameters adopted for the sensitivity to the continuum flux. The result is shown in Fig.~\ref{f:line-sens}

%
%

As can be seen from Figs.~\ref{f:cont-sens} and 
\ref{f:line-sens}, Ge(111) is preferred, as it requires a lower outer radius and it allows a better sensitivity to be achieved.

\section{Relevant astrophysical open issues that can be addressed with the proposed Laue lens}

The expected sensitivity of the lens we have sketched above, assuming bent crystals of Ge(111), is unprecedented. Nowadays the best comparison can be done with the continuum sensitivity (about $10^{-6}$~photons~cm$^{-2}$~s$^{-1}$~keV$^{-1}$ in 10$^6$~s) of the INTEGRAL/ISGRI coded-mask telescope \cite{Lebrun03} and with the emission line sensitivity ( $5 \times 10^{-5}$~photons~cm$^{-2}$~s$^{-1}$ in 10$^6$~s) of the INTEGRAL/SPI coded mask telescope \cite{Roques03}. 
For comparison, we note that the lens continuum and line sensitivity is at least a factor 300 better.

\begin{figure}
\begin{center}
  \includegraphics[width=0.37\textwidth]{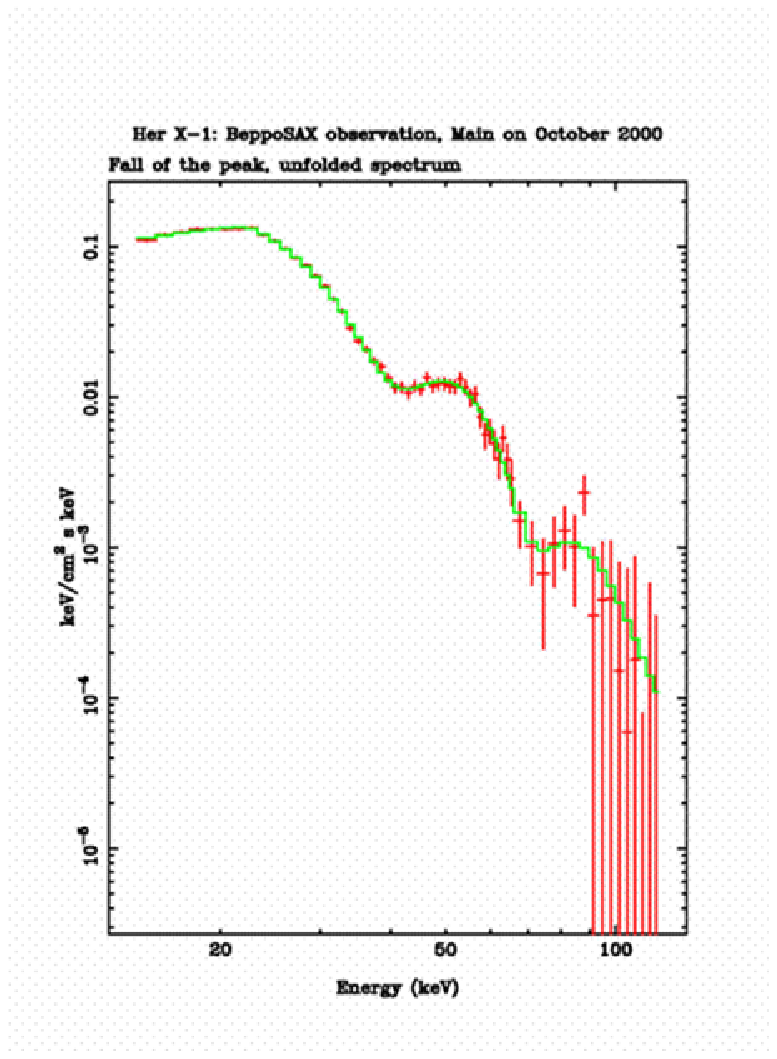}
  \includegraphics[width=0.37\textwidth]{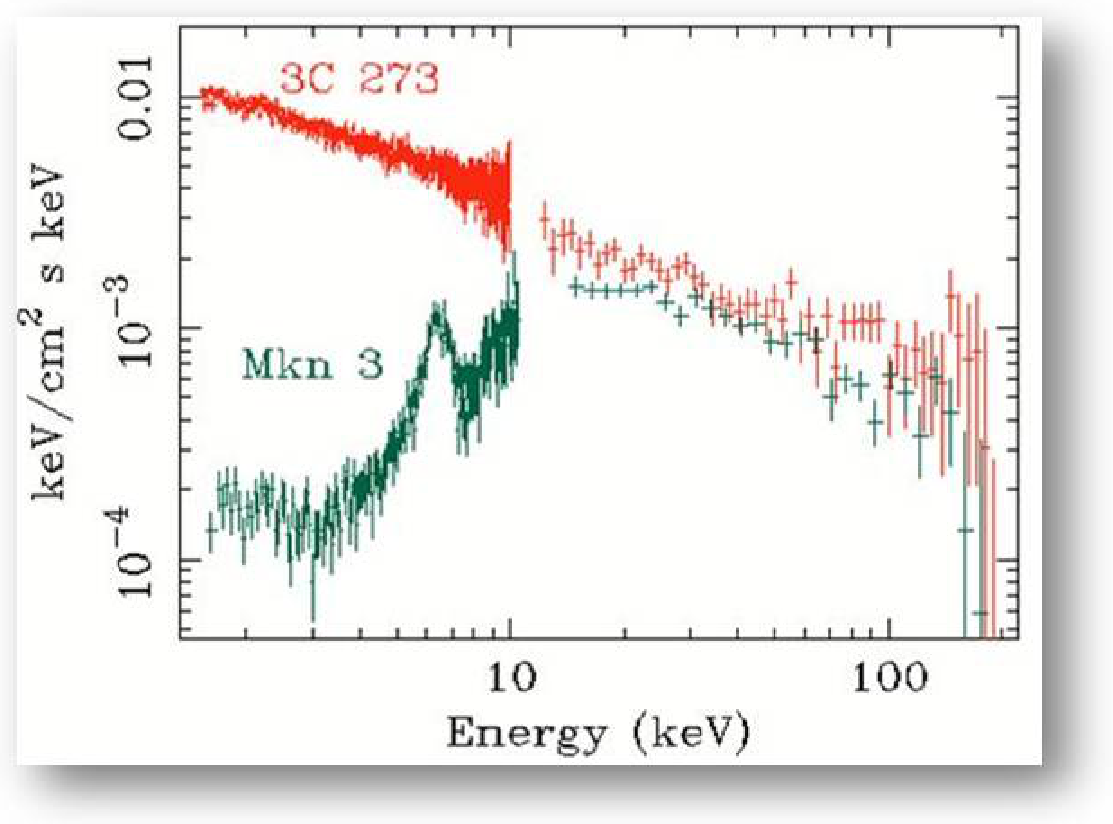}
\end{center}
\caption{The broad band measured spectra of the Galactic 
X--ray pulsar Her X--1 ({\em left}) and
of the extragalactic Active Galactic Nuclei (AGN) MKN--3 (Seyfert 2 galaxy) and 3C273
(quasar). In the case of Her X--1, the spectrum is well determined only up to 60 keV.}
\label{f:spectra}      
\end{figure}

With this sensitivity figure, the source continuum spectra beyond 100 keV can be well determined. Indeed, while below 100 keV, the spectra of the brightest sources are well determined and, in the future, those of fainter sources will be determined with \nustar, above 100 keV very little is known on the spectral energy distribution of Galactic and extragalactic sources (see two examples in Fig.~\ref{f:spectra}). With the proposed Laue lens we expect to open a new energy window and deal with many still open issues in addition to have the possibility of making unexpected discoveries.

Here we will mention some of the open issues that can be solved only with these unprecedented observations beyond 100 keV.

%
%

\subsection{Physics of accretion onto Galactic compact objects in binary systems}

The spectra of the Galactic compact sources (White Dwarfs, Neutron Stars or Black Holes) in binary systems extend from soft X--rays to beyond 100 keV. However, their knowledge beyond 100 keV (or less, see, e.g., Fig.~\ref{f:spectra}) requires much more sensitivity than that achieved  by satellite missions like \sax, \xte, and \integral. The results obtained so far clearly show that only a broad energy band that extends beyond 100 keV allows to establish the main physical components in the spectra and to study their origin. For example, in the case of X--ray pulsars, either high energy cyclotron scattering features or harmonics of lower energy features can be discovered, and thus the magnetic field strength and its properties can be investigated. 

%
One of the most interesting and debated topics is the origin of the transient powerlaw-like X-ray tail which was observed above 30 keV in some low-magnetic field NS low-mass X-ray binaries (Z and atoll bursting sources, transient sources).
Satellites which observed this feature (BeppoSAX, RXTE, Suzaku, Integral) \cite{disalvo2002, dai2007, sakurai2012, maiolino2013}, revealed only the "on" and "off" states, but nothing could be said about possible variations of the powerlaw spectral index and, most importantly, the position of the high-energy cutoff \cite{paizis2006}.
At present, models such as hybrid non-thermal Comptonization \cite{farinelli2005}, bulk motion Comptonization 
\cite{farinelli2008}
or even synchrotron emission \cite{markoff2001} can adequately fit the
data.
Only with a Laue lens telescope will be possible to thoroughly observe
the hard tail evolution and find the position of the high-energy cut-off
which would set unambigous constraints on the physical mechanism responsible for the emission.

\begin{figure}[!t]
	\begin{center}
	\includegraphics[width=0.37\textwidth]{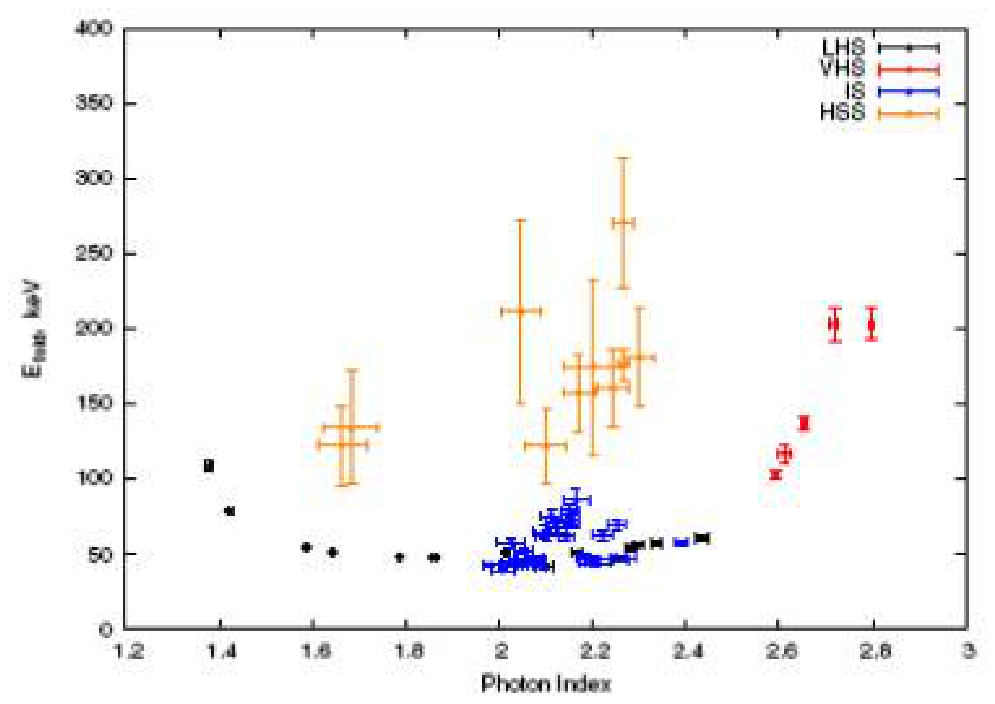}
	\includegraphics[width=0.37\textwidth]{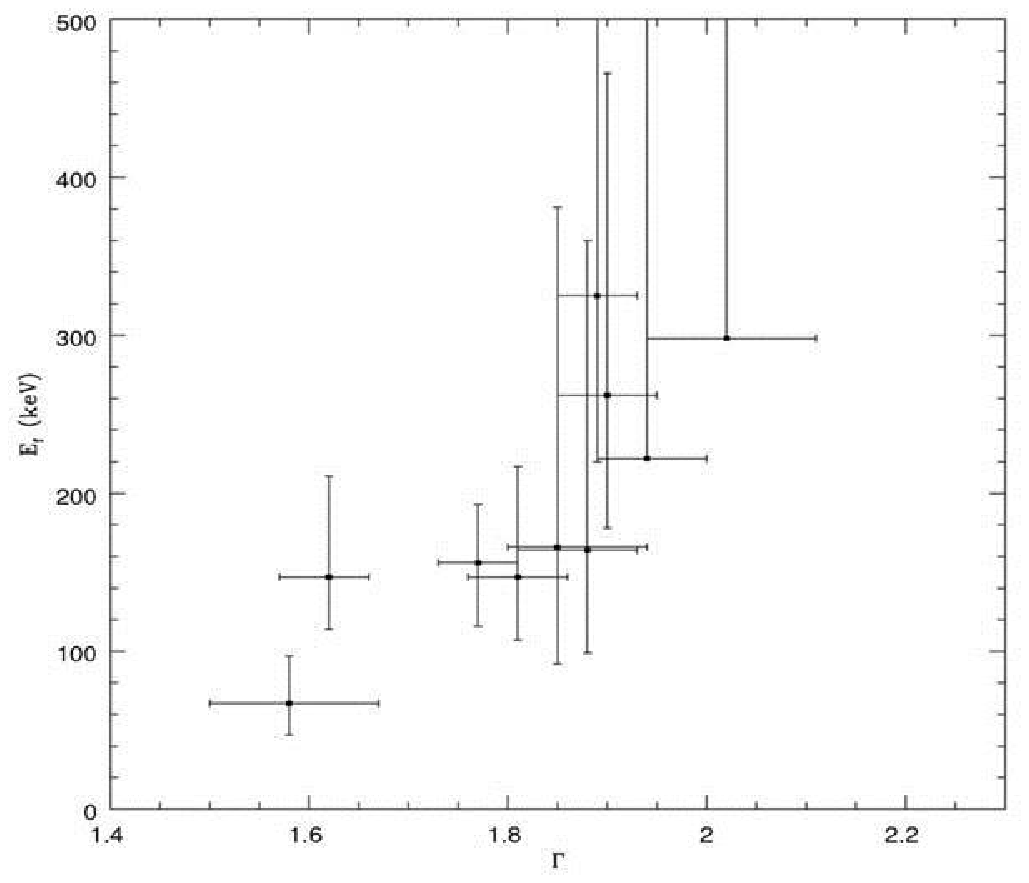}

	\end{center}
\caption{{\em Left panel}: Dependence the cutoff energy versus the photon index for the black hole transient source XTE J1550$-$564. Reprinted from Ref.~\citenum{Titarchuk10}.
{\em Riglt panel}: Dependence of the cutoff energy versus the photon index for a sample of Seyfert 1 (radio-quiet AGNs)  observed with \sax. Reprinted from Ref.~\citenum{Perola02}.}
\label{f:cutoff}
\end{figure}

The physical mechanisms responsible for the production of non-thermal emission in accreting BHs
has been demonstrated to reside in the observational appearances of the power-law tails in the X-ray spectra from these objects and their behavior in different spectral states  vs. corresponding photon index \cite{Titarchuk10,Laurent11b}.
The cutoff energies, which  can be currently measured at most during the outburst peak of the strongest BH transient sources (see, e.g., Fig.~\ref{f:cutoff}), is generally below the sensitivity limits or outside the passband of the current instrumentation, in the case of weak sources. Only the use of Laue lenses will fully address  this issue for a complete sample of BH sources.

%
%

It has been argued \cite{Laurent12} that, very close to the event horizon, X-ray photons may be upscattered by bulk electrons to MeV energies. Most of these photons fall down then in the black hole, but some of them have time to interact
anyway with another X-ray photon due to the photon-photon process, thus making an electron-positron pair. This pair creation process close to horizon can give rise to a gravitationally or not redshifted positron annihilation line. The study of this line can be carried out only through focusing instrumentation with a passband up to 600 keV.

\subsection{Magnetar physics}

Both Soft Gamma--Ray Repeaters (SGR) and Anomalous X--ray Pulsars (AXP) are now well established members of the same class of objects known as magnetars, i.e., 
neutron stars with super-strong magnetic fields ($B = 10^{14}$--$10^{16}$~G) (see review in Ref.~\citenum{Hurley10}). 
SGRs were known at high energies from many years during their burst--like activity and giant flares 
(e.g., Ref.~\citenum{Guidorzi04}), while AXPs were known to be persistent sources with pulsar peridiocities in the 2--8 s range 
and very soft X-ray ($\le 10$~keV) spectra (blackbody plus a power-law  with index between 2 and 4). 
Nowadays we know that also SGRs show similar periodicities  during their quiescence states and that AXPs show burst-like activity. 
However the softness of the AXP spectra did not  predict detections at energies above 10 keV. So it was with a great surprise 
that the imaging instrument IBIS aboard the INTEGRAL satellite measured hard X-rays 
(see Ref.~\citenum{Gotz06,Kuiper06} and references therein). 

\begin{figure}[!t]
	\begin{center}
	\includegraphics[width=0.4\textwidth]{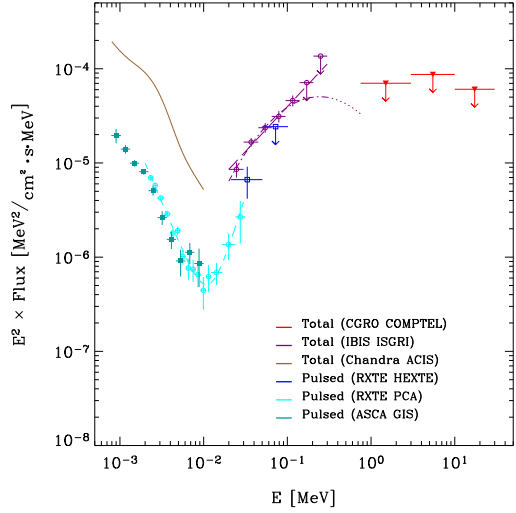}
	\end{center}
\caption{Example of the status of our 
knowledge of the X--/gamma--ray spectra of AXPs and SGRs: the case of the AXP 4U~0142+61. 
Reprinted from Ref.~\cite{Kuiper06}.}
\label{f:SGR-AXP}
\end{figure}

Broad band spectra of the persistent emission from these sources beyond 10 keV can be described with a power law with photon index less than 2, which would imply a 
divergent energy output in the case of no spectral cutoff. However, the cutoff energies have not been 
measured yet (see an example in 
Fig.~\ref{f:SGR-AXP}). Thus the question of the physical origin of the high energy component ($>$100 keV) in magnetars, for which there are several models (see, e.g., Ref.~\citenum{Thompson05}), can be solved only with the focusing telescope of the type proposed here. 
Indeed from the lens sensitivity in 10$^5$~s shown in
Fig.~\ref{f:cont-sens}, we get, in the units shown in 
Fig.~\ref{f:SGR-AXP}, a limiting sensitivity of about $2\times 10^{-7}$~MeV~cm$^{-2}$~s$^{-1}$~MeV$^{-1}$, which is about 3 order of magnitude lower than the upper limit shown in the figure.  

%
%

\subsection{Accretion physics in Active Galactic Nuclei (AGN)}

In the case of radio--quiet AGNs, like  Seyfert galaxies, high energy observations allow one to probe the inner regions of the accretion disk and, in particular, the properties of the corona around the Massive Black Hole (MBH). While there is general consensus that the X--ray emission is due to Comptonization of UV/soft X--ray disk photons off this hot corona, the coronal electron temperature and its optical depth are poorly known, given that most of them can be observed beyond 100 keV, where the sensitivity of the current instrumentation is background limited (see, e.,g. Fig.~\ref{f:spectra}). The most sensitive attempts were done with \sax\ (see 
Ref.~\citenum{Perola02}) with the results shown in 
Fig.~\ref{f:cutoff} right panel. 
More recently, broad band observations with INTEGRAL and a variety of soft X-ray  telescopes have made another step in the measurement
of high energy cut-offs. Even with all the limitation given by this type of  analysis (spectral complexity and non simultaneity of the soft and hard X-ray
data), a number of cut-offs have been measured and lower limits reported for complete sample of AGNs. The distribution of the measured cut off energies clusters 
around 100 keV, while the bulk of the
lower limits on this parameter are found below 300 keV 
\cite{Bassani13}. 
One must then conclude that the  high energy
cut-off is present and that  only the joint use of focusing telescopes, like those aboard \nustar, combined with a  Laue lens to probe the energy range above 100 keV, 
can solve this issue.

\begin{figure}[!t]
	\begin{center}
	\includegraphics[width=0.37\textwidth]{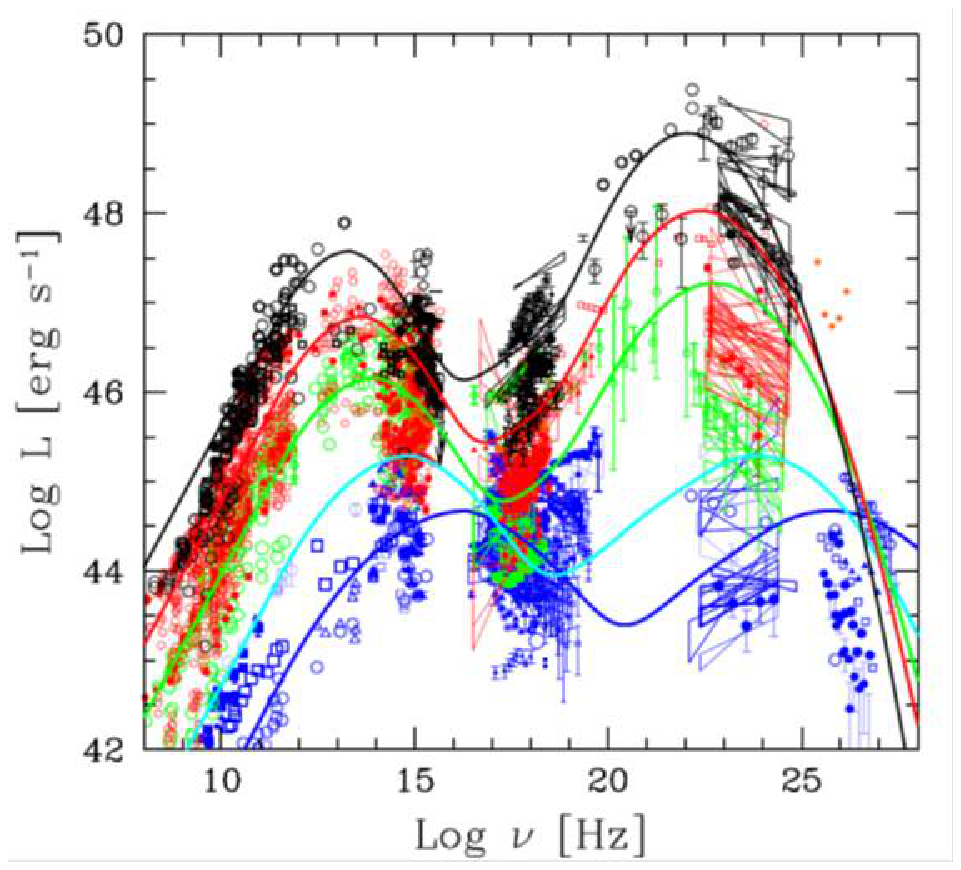}
	\includegraphics[width=0.37\textwidth]{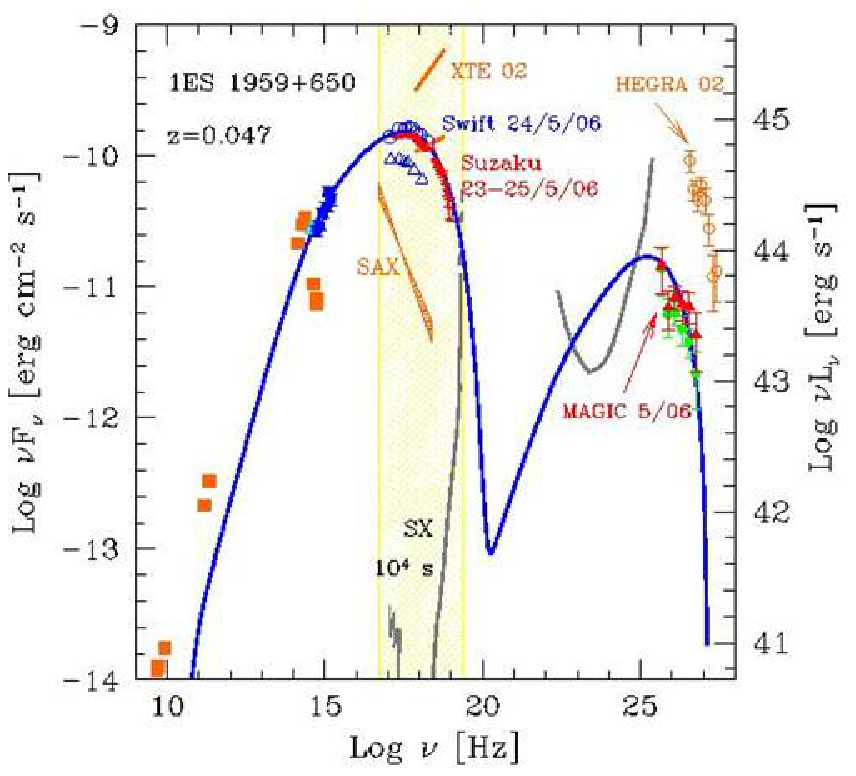}
	\end{center}
\caption{{\em Left panel}: Broad band spectra of blazars of increasing luminosity. Reprinted from Ref.~\citenum{Ghisellini11b}.
{\em Riglt panel}: Example of a blazar source (the BL-Lac source 1ES 1959+650)that shows its expected dip at about 400 keV, that could be detected with the Laue lens we are proposing.  Reprinted from Ref.~\citenum{Ghisellini09a}.}
\label{f:blazar}
\end{figure}

In the case of radio-loud AGNs (Blazars), with prominent jets, the spectra show two main humps, one peaking from millimeter to the X--ray spectral range, the other from the hard X--ray to the high energy gamma--ray range. The first peak is is interpreted as due to synchrotron emission, while the second one as due to Inverse Compton. The highest luminosity Blazars are Flat Spectrum Radio Quasars (FSRQ) and show their Compton peak in the passband of Laue lenses (see left panel of Fig.~\ref{f:blazar}), while BL-Lac blazars show, in the energy passband of Laue lenses, the expected dip between the two humps.  Only with a sensitive instrument beyond 100 keV, like the Laue lens we are proposing here, we can detect the expected  dip between humps and confirm the current models.

%
%

\subsection{Origin of the Cosmic X--ray Background}

\begin{figure}[!t]
	\begin{center}
	\includegraphics[width=0.4\textwidth]{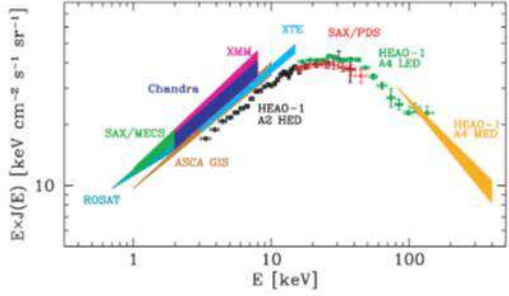}
	\end{center}
\caption{EF(E) spectrum of the Cosmic X--ray Background. Reprinted from Ref.~\citenum{Frontera07b}.}
\label{f:cxb}
\end{figure}

The Cosmic X--ray background (CXB) is characterized by a $E F(E)$ spectrum with a peak around 30 keV followed by a well defined decrease beyond this energy (see Fig.~\ref{f:cxb}). Currently, in the X-ray band the CXB has been substantially resolved in terms of obscured and unobscured AGNs (e.g., Ref.~\citenum{Xue11}), while, around 30 keV, the background composition is being investigated with \nustar\ 
\cite{Alexander13}.

%
%

Instead, the origin of the slope of the spectrum above 30 keV is still debated. CXB synthesis models \cite{Gilli07} assume a combination of unobscured, Compton thin and Compton thick radio-quiet AGN 
populations with different photon index distributions and fixed high energy spectral cutoff $E_c$.
However we know from the observation of the strongest AGNs that $E_c$ changes from an AGN to another. Thus it is not realistic to assume a fixed $E_c$. 
In addition to radio-quiet AGNs, a contribution to the high energy part of CXB is certainly due to blazars, their real contribution is still matter of discussion due to the absence of deep observations beyond 70--100 keV, where they show their maximum power \cite{Ghisellini09a}.
 The best spectral studies of blazars can be done only extending the energy band of focusing instruments like those aboard \nustar\ beyond 100 keV. This can only be obtained by including Laue lenses in future missions.

\subsection{Positron astrophysics}

\begin{figure}[!t]
	\begin{center}
	\includegraphics[width=0.4\textwidth]{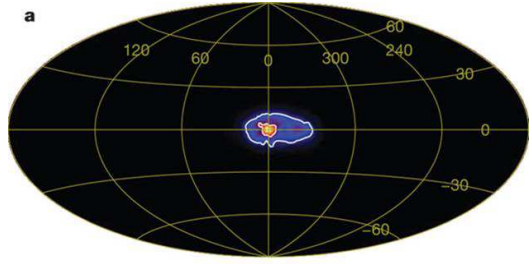}
	\end{center}
\caption{Distribution in Galactic coordinates of the positron-annihilation line. Reprinted from Ref.~\citenum{Weidenspointner08}.}
\label{f:ann-line}
\end{figure}

Positron production occurs in a variety of cosmic explosions and acceleration sites, and the observation 
of the characteristic 511 keV annihilation line provides a powerful tool to probe plasma 
composition, temperature, density and ionization degree. The positron annihilation signature is 
readily observed from the Galactic bulge region, but yet the origin of the positrons remains mysterious. A SPI/INTEGRAL all-sky map \cite{Weidenspointner08} of galactic 
e$^-$e$^+$ annihilation radiation shows an asymmetric distribution of 511 keV emission with a total flux of $1 \times 10^{-3}$~photons~cm$^{-2}$~s$^{-1}$ (see Fig.~\ref{f:ann-line}), however its origin is still a mystery. The annihilation line  can be due to the integrated contribution of positron annihilation features from low mass X-ray binaries with strong emission at photon energies $>$20 keV (hard LMXBs), as suggested by the authors of the \integral\ discovery, but it could also have a different origin. For example, it could be due to the presence of antimatter, or to dark matter annihilation as proposed in  Ref.~\citenum{Finkbeiner07}),
or to the existence of a source of radioactive 
elements, like $^{26}$Al, $^{56}$Co, $^{44}$Ti, or finally to the presence of a gamma--ray source (e.g., gamma--ray pulsar).
 
%
%

To unveil the mystery about the origin of the 511 keV line from the Galactic Center Region, much more sensitivity and angular resolution is needed, achievable only with Laue lenses. In the lens configuration that we propose in this paper, the sensitivity to a positron annihilation line is 4 orders of magnitude better than the observed line flux.

\subsection{Indirect detection of dark matter} 

Understanding the nature of Dark Matter (DM), which makes up almost a quarter of the mass-energy density of the Universe, is one of the most pressing open issues of fundamental physics today. Indirect detection of DM is generally referred to as all those techniques to observe radiation and particles which are the products of DM annihilations ($\gamma$ rays, neutrinos, high-energy electrons and positrons). 
The flux of such radiation is proportional to the annihilation rate and therefore to the square of the DM density for a given DM annihilation cross section $\sigma$, i.e. $\propto \sigma v\, n^2(r)=\sigma v\rho_{\rm DM}^2(r)/m^2_{\rm DM}$. 
As a result, the most promising sites for such searches are those with high DM density $\rho_{\rm DM}$, such as the galactic center, nearby spheroidal dwarf galaxies and massive galaxy clusters. In all these systems the DM density profile is expected from simulations, and constrained by observations, to increase as a power-law, $\rho_{\rm DM}\propto r^{-\alpha}$, with $\alpha\lesssim 1$. Several attempts to detect a $\gamma$-ray excess from Milky Way satellites have been made with 
\fermi\ at 0.1-100 GeV (e.g. Ref.~\citenum{Ackermann11}), which have provided new upper limits. 
It is clear that the two order of magnitude increase in sensitivity and angular resolution afforded by focusing Laue lenses, with good sensitivity up to 600 keV, could lead to a breakthrough of this field, with an uncontroversial detection of a $\gamma$-ray flux as final product of DM annihilation chain, or alternatively by setting very low upper limits which will strongly constrain DM models.

\subsection{Unveiling the GRB hard X-ray afterglow emission}

Despite the enormous progress occurred in the last 20 years, the Gamma-Ray Bursts (GRB) phenomenon is still far to be fully understood. One of the most important open issues is the afterglow emission above 10 keV, which is almost completely unexplored, as a result of the lack of sensitive of detectors operating in this energy band. The only detection of the hard X-ray emission from a GRB (the very bright GRB 990123) performed with the BeppoSAX/PDS instrument (15-200 keV), combined with optical and radio observations, have seriously challenged the standard scenario. In the latter the dominant mechanism is synchrotron radiation produced in the shock of an 
ultra-relativistic fireball with the ISM, showing the need of a substantial revision of current models 
\cite{Corsi05,Maiorano05}. The possibility to focus hard X--rays up to several hundreds of keV can provide an important breakthrough in this field, through follow-up observations of bright GRBs detected and localized by GRB dedicated experiments. 

\subsection{High sensitivity polarization measurements}

Several classes of sources (e.g., NS and BH in binary systems, AGNs, GRBs) are expected to show hard X--ray polarization. The last discovery of a polarized component in the Cygnus X--1 spectrum above 400 keV confirms the expectations. With Laue lenses, thanks to the focusing capability, with a proper focal plane detector, the possibility of discovering new polarized components from celestial sources becomes solid. We expect exciting results in this field.

\section{Conclusions}

We have reported on the conclusion of the first--phase of the LAUE project, specifically the new capability we developed to accurately build Laue lens petals with a focal length of 20 m, with the  production of bent crystals of Ge(111) and GaAs(220). The tests of these crystals give reflectivity results consistent with the expectations. On the basis of the expected PSF and reasonable assumptions on the background in orbit, we evaluated the expected sensitivity to both continuum emission and to narrow emission lines in the passband (90--600 keV) of a  lens based on petals like the one we are assembling.  The expected sensitivity is about 300 times better than the current instrumentation, opening a new window in hard X--ray astronomy. We have discussed possible astrophysical open issues that can be tackled, showing that for the first time we can achieve unprecedented goals and  make unexpected discoveries.

\acknowledgments     

The LAUE project is the result of big efforts made by a large number of organizations and people.
We would like to thank all of them. We also acknowledge the ASI Italian Space Agency for its support 
to the LAUE project under contract I/068/09/0.

Vincenzo Liccardo and Vineeth Valsan are supported by the Erasmus Mundus Joint Doctorate Program by Grant Number
2010-1816 from the EACEA of the European Commission.

\bibliography{lens_biblio}
\bibliographystyle{spiebib}

\end{document}